\documentclass[conference]{IEEEtran}

\usepackage{amsmath}
\usepackage{amssymb}
\usepackage{graphicx}
\usepackage{subcaption}

\usepackage{algorithm}
\usepackage{algpseudocode}

\usepackage{booktabs}
\usepackage{threeparttable}
\usepackage{multirow}
\usepackage{tabularx}
\usepackage{array}

\usepackage[table]{xcolor}

\usepackage{pdflscape}

\usepackage{pifont}

\usepackage{tikz}
\usepackage{pgfplots}
\usepgfplotslibrary{groupplots,statistics}
\usetikzlibrary{calc}
\pgfplotsset{compat=1.18}

\begin{document}

\title{Bridging the Semantic Gap from Natural Language Specifications to RTL with Layered Formal Guidance}

\author{\IEEEauthorblockN{Yan Tan, Xiangchen Meng and Yangdi Lyu$^{\dagger}$}
\IEEEauthorblockA{Microelectronics Thrust, The Hong Kong University of Science and Technology (Guangzhou) \\
$^{\dagger}$Corresponding author: yangdilyu@hkust-gz.edu.cn}
}
\maketitle

\begin{abstract}
While large language models (LLMs) have shown promise in generating register-transfer level (RTL) code from natural-language specifications, their reliability remains limited on designs with complex control logic and temporal behavior. This stems from a fundamental semantic gap: high-level natural language often leaves implementation-critical details underspecified, forcing LLMs to infer unstated assumptions. This paper presents Layered Formal Guidance, a novel framework that bridges this semantic gap by using formal properties as an intermediate semantic layer between natural-language specifications and RTL generation. Our framework applies formal guidance across three layers: property-driven specification augmentation before generation, property-consistency checking after generation, and counterexample-guided refinement after checking. Experiments on RTLLM benchmarks and various open-source IP designs demonstrate that our framework achieves higher functional correctness and lower sampled output-trace diversity. Furthermore, we show that Layered Formal Guidance can flexibly augment both general-purpose LLMs and domain-specific models to substantially improve their generation reliability.
\end{abstract}

\begin{IEEEkeywords}
LLM-generated Verilog, specification refinement, formal guidance,
RTL generation
\end{IEEEkeywords}

\section{Introduction}
\label{sec:introduction}

Existing studies show that LLMs can successfully generate syntactically valid Verilog for simple modules~\cite{Benchmark_RTL_2022, Dehaerne_Verilog_2023, OpenLLM, VerilogEval}. However, LLMs still struggle to produce functionally correct designs when the logic is complex or the Verilog code is long. A key challenge is the semantic gap between high-level natural-language (NL) intent and implementation-critical RTL details. While natural language specifications usually describe high-level functional intent, RTL must rigidly define hardware behavior across every clock cycle, state, and input condition. Consequently, LLMs struggle to bridge high-level intent with implementation-critical RTL details, such as cycle-accurate timing constraints, state updates, interface protocols, and corner-case behavior.

This semantic abstraction gap makes NL-to-RTL generation more than a simple code translation task. A specification may appear clear to a human designer, but it can still leave important RTL behaviors undefined. 
For example, a high-level divider specification may define only the mathematical relationship among the dividend, quotient, and remainder. However, it may omit corner-case behaviors such as division by zero, interface protocols like ready-valid handshakes, or exact timing constraints and output validity windows. When generating RTL, the LLM must infer these unstated semantics, causing divergent assumptions to produce inconsistent or buggy RTL implementations from the same description.

Existing agentic and feedback-driven methods improve generated RTL using compiler messages, simulation logs, or iterative repair~\cite{verilogcoder,Xie_2023,mage,vflow}. However, these methods mainly correct implementation errors in an already generated RTL candidate. When the underlying natural-language specification is incomplete, ambiguous, or underspecified, repairing one RTL candidate cannot resolve the underlying semantic uncertainty. Therefore, post-generation RTL repair alone is insufficient to bridge the semantic abstraction gap and capture key behavioral constraints.

To bridge this gap, we use SystemVerilog Assertions (SVAs) as an intermediate layer to formalize candidate cycle-level rules for timing, state updates, and interface behavior. Before RTL generation, the framework derives these properties from the specification, translates them into structured natural-language constraints, and provides both the original specification and these explicit constraints to the generator. This exposes unspecified behaviors and restricts semantic freedom. As inconsistencies may still exist among the specification, the generated properties, and the RTL, the framework model-checks each RTL candidate against the current SVA set. When an SVA property is violated, the framework analyzes the counterexample (CEX) trace returned by the formal checker. It then determines whether the inconsistency is due to an RTL error, an omission or ambiguity in the specification, or an incorrect SVA property. The framework then revises the corresponding artifact. This in-loop check only evaluates consistency against the SVA set, while final functional correctness is independently verified using a fixed golden reference.

Our main contributions are summarized as follows:

\begin{itemize}
    \item We propose Layered Formal Guidance, a framework that bridges natural-language specifications and RTL generation by applying formal guidance across specification augmentation, candidate model checking, and iterative failure attribution.
    \item We develop a property-driven specification augmentation mechanism that exposes implicit RTL-level requirements and restricts semantic freedom prior to code generation.
    \item We introduce a counterexample-guided failure attribution and refinement mechanism that maps formal failures to targeted refinements, including RTL repair, specification disambiguation, and property correction.
    \item We empirically show that the proposed framework improves RTL generation correctness, semantic consistency, and reliability on various benchmarks and IP-derived designs.
\end{itemize}

\section{Related Work}

\subsection{LLM-Based RTL Generation}

Recent work has explored LLM-based RTL generation through benchmarks, domain-specific training, and inference-time refinement. Benchmarks such as VerilogEval~\cite{VerilogEval} and OpenLLM-RTL~\cite{OpenLLM} provide evaluation tasks and datasets for measuring Verilog generation quality. Training- and data-centric methods, such as RTLCoder~\cite{Xie_2023}, OriGen~\cite{2024origen}, BetterV~\cite{BetterV}, ChipSeek-R1~\cite{chen2025chipseek}, and CodeV-R1~\cite{zhuqimeng}, improve RTL generation through hardware-specific corpora, augmentation, self-reflection, or reward-based optimization.

Parallel to generator-focused techniques, several methods introduce feedback-driven refinement during inference. AutoChip~\cite{AutoChip} uses compiler and tool feedback for iterative HDL generation, while RTLFixer~\cite{RTLFixer} performs retrieval-assisted debugging for RTL repair. In addition to debug information, high-level programming languages are also used to fix RTL by generating reference models and testbenches to guide functional correction~\cite{11420300}. However, these feedback loops operate strictly after code generation, leaving the underlying specification ambiguities untouched.

\subsection{Formal Methods for LLM-Based Hardware Generation}

Prior work has used formal methods in LLM-based RTL generation mainly for assertion generation and post-generation verification. AssertLLM~\cite{yan2025assertllm} and Spec2Assertion~\cite{wu2025spec2assertion}  generate assertions from natural-language specifications. AssertionForge~\cite{bai2025assertionforge} combines information from the specification and RTL to improve assertion generation. These studies show that LLMs can translate informal intent into checkable properties. 

On the verification side, FormalRTL~\cite{li2026formalrtl} uses an executable C reference model and formal equivalence checking to verify and refine generated RTL. VeriSure~\cite{liu2026verisure} combines trace analysis, assertion checking, and Boolean equivalence to find and repair errors in specific parts of the RTL. These works mainly use formal methods for assertion generation or post-generation checking. In contrast, our framework integrates formal properties across all generation stages, resolving specification ambiguities before code generation and leveraging formal verification failures to jointly refine the RTL, specification, and property set.

\section{Methodology}

\subsection{Motivation}

Natural language prompts typically describe high-level functional intent while leaving essential implementation requirements implicit. When generating RTL, the LLM is forced to independently infer these unstated implementation details, such as cycle-accurate timing, state update mechanisms, control logic behavior, and structural partitioning. As a result, repeated sampling from an underspecified prompt produces multiple syntactically valid RTL candidates that appear prompt-consistent but exhibit severe functional divergence during simulation. 

\begin{figure}[t]
    \centering
    \includegraphics[width=1\linewidth]{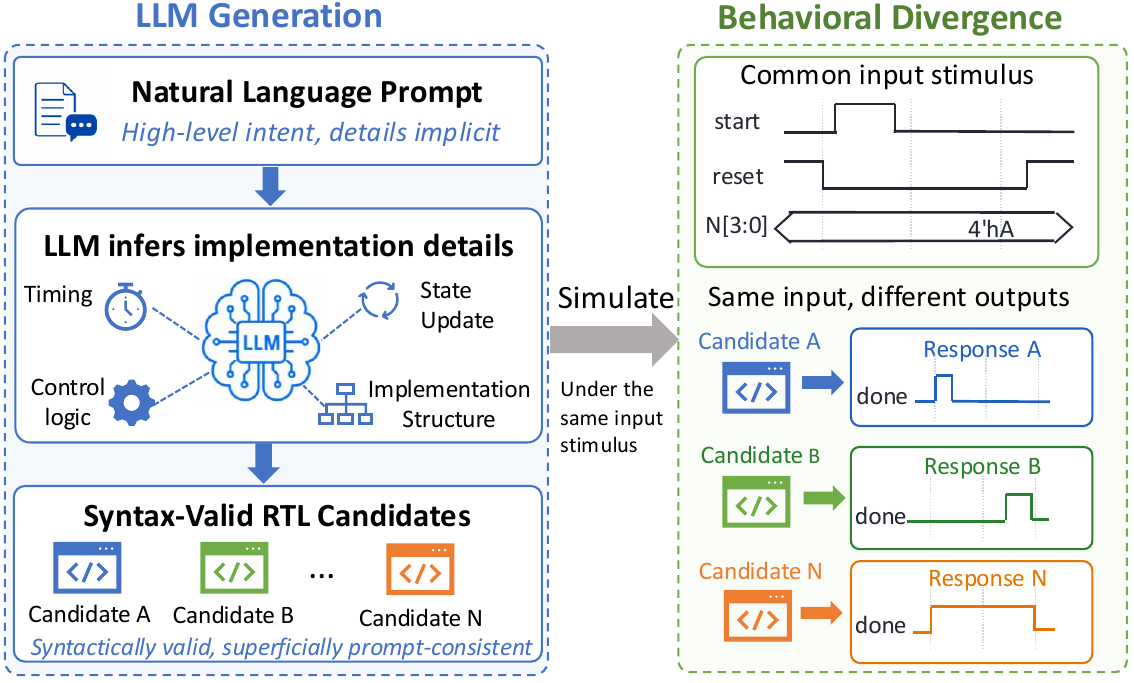}
    \caption{Underspecified prompts can lead to syntactically valid but behaviorally divergent RTL candidates.}
    \label{fig:Gap}
\end{figure}

An example of this divergence is illustrated in Fig.~\ref{fig:Gap}. Driven by the same input stimulus, different generated candidates yield conflicting output trace responses. Because this behavioral divergence stems from fundamental specification ambiguity rather than simple syntax bugs, post-generation RTL repair alone cannot resolve the underlying uncertainty. The generation framework must therefore formalize unstated cycle-level rules to narrow the semantic space before code generation begins.

\begin{figure*}[t]
    \centering
    \includegraphics[width=1\linewidth]{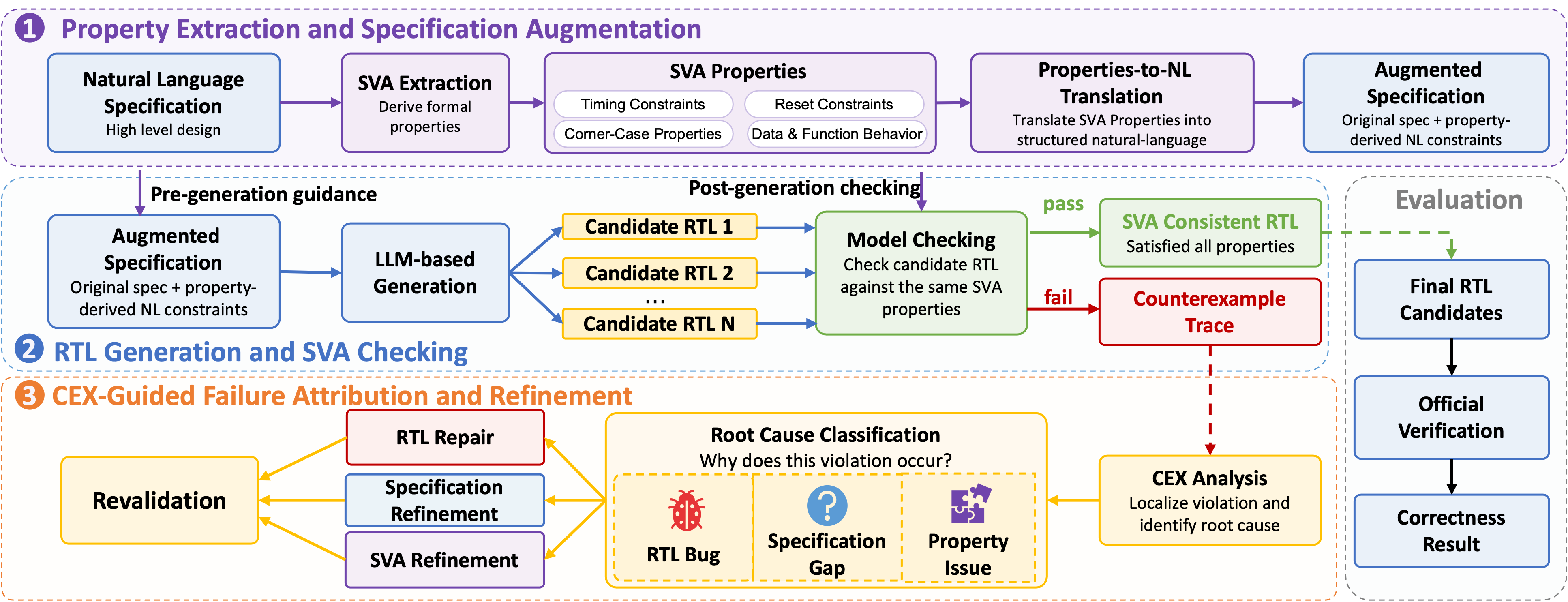}
    \caption{Overview of the Layered Formal Guidance framework. }
    \label{fig:overview}
\end{figure*}

\subsection{Overview}
Fig.~\ref{fig:overview} illustrates the overall workflow of Layered Formal Guidance. The framework is organized into three layers. Layer 1 derives candidate SVA properties from the original specification and translates them into structured constraints to augment the prompt. Layer 2 generates candidate RTLs from the augmented specification and evaluates each candidate against the SVA properties. When a property fails, the formal tool produces a counterexample trace. Layer 3 uses this trace to attribute the failure and perform targeted refinement across the RTL, the specification, or the property set. Depending on the diagnosis, the framework either directly re-verifies a repaired RTL candidate or regenerates the RTL from updated specifications and SVA properties. \textit{SVA checking operates within this generation and refinement loop, while final functional correctness is evaluated separately using an independent golden reference.}

\subsection{Property Extraction and Specification Augmentation}

In Layer 1, the framework extracts an initial set of SVA properties from the natural language specification. Each property formalizes a critical behavior that must be defined at the RTL level, such as reset logic, temporal constraints, state transitions, interface handshakes, and corner cases, thereby supplying the LLM with explicit implementation requirements.

\begin{figure*}[t]
    \centering
    \includegraphics[width=1\linewidth]{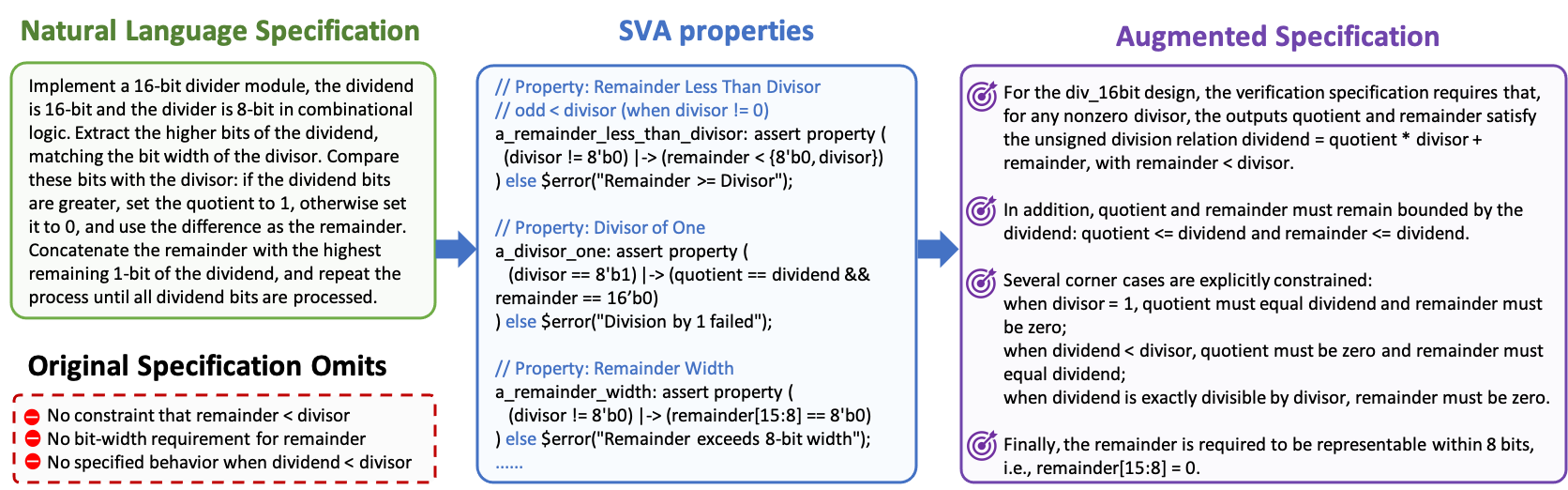}
    \caption{Formal-property-derived specification augmentation for an unsigned divider.}
    \label{fig:sva_extraction}
\end{figure*}

After basic checking, we translate the retained SVA properties into concise natural-language constraints for RTL generation. The translation keeps the condition, timing relation, and required response of each property. The resulting constraints are appended to the original prompt, forming an augmented specification for RTL generation. The original specification provides the design function and interface, while the appended constraints describe more specific RTL-level requirements. Fig.~\ref{fig:sva_extraction} illustrates this process with an unsigned divider. In this example, properties on quotient-remainder consistency and division-by-zero handling are translated into natural-language constraints and appended to the original specification.

\subsection{RTL Generation and SVA Checking}

The augmented specification from Layer 1 provides more specific RTL-level requirements for RTL generation. However, the LLM generator may still make implementation errors during RTL generation. Layer 2 therefore checks whether the generated RTL satisfies these requirements. The framework passes the generated Verilog together with the SystemVerilog Assertion set to an automated model checker. If all properties pass, the current generation completes. If any property fails, the formal tool returns the failed property and a counterexample trace. We pass the failed property and counterexample to Layer 3 for failure analysis and CEX-guided refinement.

\subsection{CEX-Guided Failure Attribution and Refinement}

\begin{figure}[h]
    \centering
    \includegraphics[width=1\linewidth]{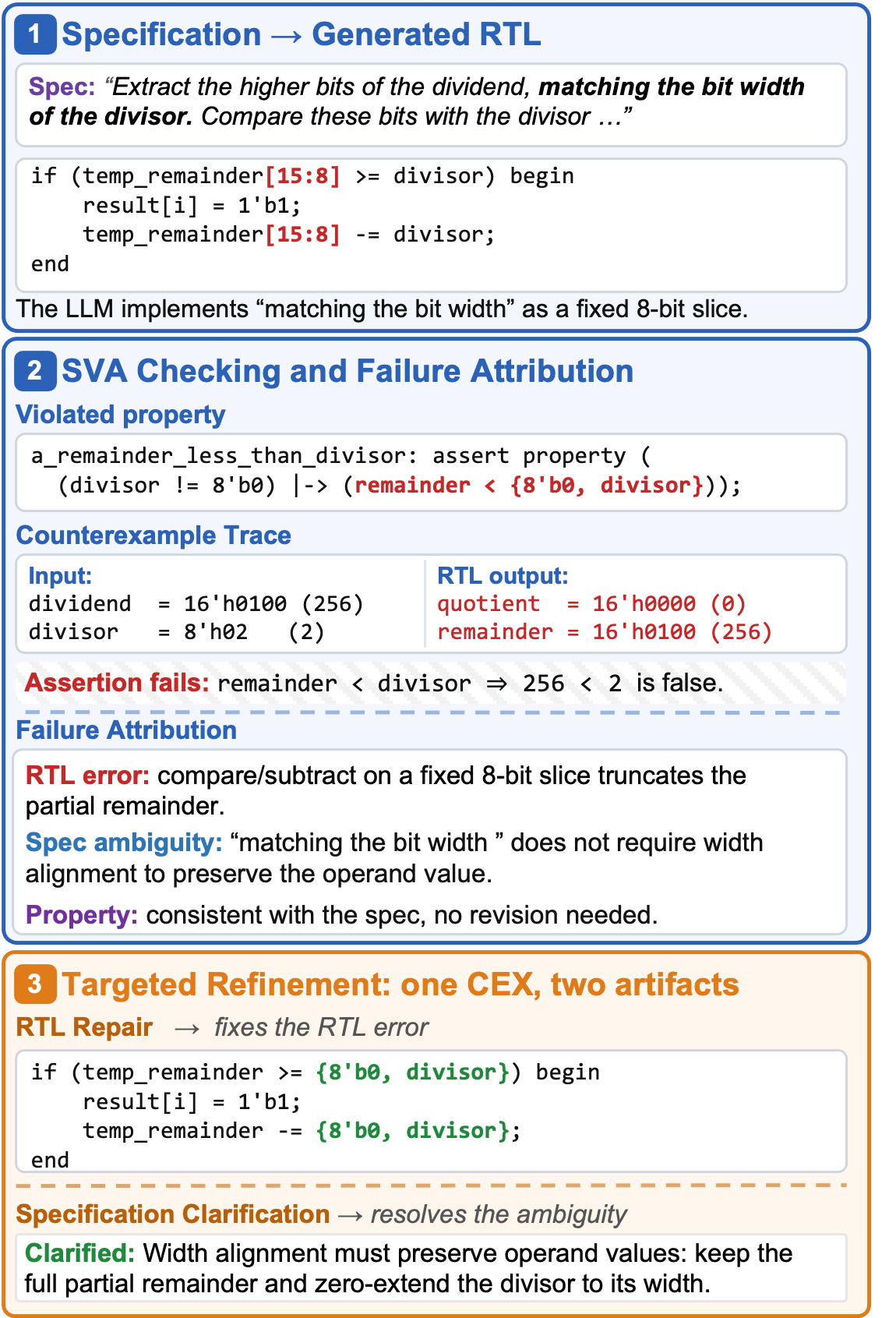}
    \caption{Example of counterexample-guided refinement for an unsigned divider.}
    \label{fig:cex}
\end{figure}

For each RTL candidate that fails SVA checking, the analyzer receives the specification, the current RTL candidate, the violated SVA property, and the counterexample trace. The counterexample provides the concrete inputs, states, and RTL behavior associated with the violation. The analyzer evaluates these diagnostic signals to isolate the failure root cause and execute targeted updates across three potential artifacts:
\begin{itemize}
    \item RTL Implementation Bugs: when the candidate RTL fails to implement a requirement already stated in the specification, the framework repairs or regenerates the candidate RTL directly.
    \item Specification Ambiguities: when the counterexample exposes RTL-level behavior that the original specification does not define sufficiently, the framework clarifies the prompt and then regenerates the RTL under the updated specification.
    \item Property Errors: when an SVA conflicts with the original specification, misses a required condition, or incorrectly restricts legal behavior, the framework revises the property alongside its translated prompt constraint and then regenerates the RTL.
\end{itemize}

If a single trace reveals flaws across multiple artifacts, the framework updates all affected components simultaneously.
Each RTL candidate is refined independently. After each refinement, the framework reruns SVA checking on the current property set. The refinement loop stops when the candidate satisfies the current properties, when further refinement makes no progress, or when the preset iteration limit is reached.

Fig.~\ref{fig:cex} illustrates this counterexample-guided refinement process using an unsigned divider module implementing an iterative long-division algorithm. In this task, during Layer 1 specification augmentation, the framework extracts a candidate formal property on operand alignment and translates it into an explicit prompt requirement to instruct the model to match operand width. Although clear to a human designer, this phrasing contains a semantic gap. During Layer 2 RTL generation, the LLM literally translates operand width requirements by slicing temp\_remainder[15:8] to match the divisor width. This slicing of the partial remainder prior to subtracting the divisor introduces functional errors and yields an incorrect result. In Layer 2 model checking, SVA checking catches this functional violation, and the counterexample provides the exact cycle-by-cycle inputs and state transitions triggering the failure. In Layer 3, based on this diagnostic feedback, the framework repairs the RTL by clarifying the corresponding width-handling requirement, and preserving the full temp\_remainder while zero-extending the narrower divisor. This example shows that while pre-generation specification augmentation can reduce underspecified behavior, it cannot guarantee that the requirements are correctly implemented in RTL. Post-generation SVA checking and CEX-guided refinement are therefore still needed.

\section{Experiments}
\label{sec:experiments}
\subsection{Experimental Setup}
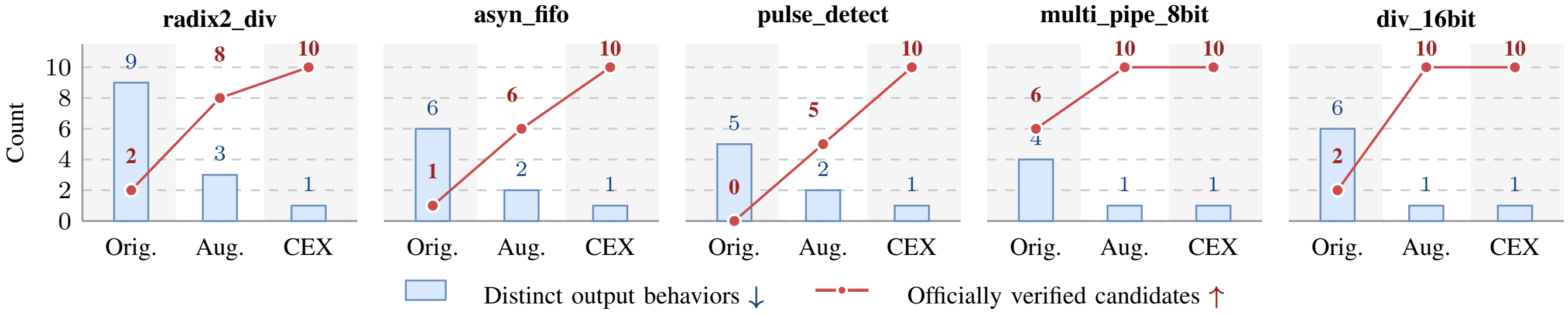
\begin{figure*}[t]
\centering
\definecolor{BarBlue}{RGB}{218, 232, 252}
\definecolor{BarEdge}{RGB}{108, 142, 191}
\definecolor{TextBlue}{RGB}{17, 65, 126}
\definecolor{LineRed}{RGB}{204, 76, 76}
\definecolor{TextRed}{RGB}{153, 35, 35}
\definecolor{BgGray}{RGB}{245, 245, 245}

\begin{tikzpicture}
\pgfplotsset{set layers}

\begin{groupplot}[
    group style={
        group size=5 by 1,
        horizontal sep=0.3cm
    },
    scale only axis,               
    width={(\textwidth-2.6cm)/5},  
    height=2cm,
    ymin=0, ymax=11.5,
    xmin=0.45, xmax=3.55,
    xtick={1,2,3},
    xticklabels={Orig., Aug., CEX},
    x tick label style={font=\footnotesize, text depth=0ex, yshift=-2pt},
    ytick={0,2,4,6,8,10},
    y tick label style={font=\footnotesize},
    tick style={draw=none},
    title style={font=\footnotesize\bfseries, yshift=-4pt},
    enlarge x limits=false,
    clip=false,
    axis x line*=bottom,
    axis y line*=left,
    ymajorgrids=true,
    grid style={dashed, gray!40, semithick},
    y axis line style={opacity=0},
    x axis line style={gray!80, semithick},
    bar style/.style={
        ybar, bar width=11pt, fill=BarBlue, draw=BarEdge, semithick, mark=none,
        nodes near coords,
        every node near coord/.append style={font=\scriptsize\bfseries, color=TextBlue, yshift=1pt}
    },
    line style/.style={
        color=LineRed, thick, solid, mark=*,
        mark options={solid, fill=LineRed, draw=white, line width=0.8pt, scale=1.0},
        nodes near coords, point meta=explicit symbolic,
        every node near coord/.append style={font=\scriptsize\bfseries, color=TextRed}
    },
]

\nextgroupplot[
    title={radix2\_div},
    ylabel={Count},
    ylabel style={font=\footnotesize},
    y axis line style={gray!80, semithick, opacity=1},
]
\begin{pgfonlayer}{axis background}
    \fill[BgGray] (axis cs:0.45,0) rectangle (axis cs:1.5,11.5);
    \fill[BgGray] (axis cs:2.5,0) rectangle (axis cs:3.55,11.5);
\end{pgfonlayer}
\addplot+[bar style] coordinates {(1,9) (2,3) (3,1)};
\addplot+[line style] coordinates {
    (1,2)  [\raisebox{5pt}{2}]
    (2,8)  [\raisebox{8pt}{8}]
    (3,10) [\raisebox{-13pt}{10}]
};

\nextgroupplot[title={asyn\_fifo}, yticklabels=\empty]
\begin{pgfonlayer}{axis background}
    \fill[BgGray] (axis cs:0.45,0) rectangle (axis cs:1.5,11.5);
    \fill[BgGray] (axis cs:2.5,0) rectangle (axis cs:3.55,11.5);
\end{pgfonlayer}
\addplot+[bar style] coordinates {(1,6) (2,2) (3,1)};
\addplot+[line style] coordinates {
    (1,1)  [\raisebox{5pt}{1}]
    (2,6)  [\raisebox{5pt}{\hspace{-6pt}6}]
    (3,10) [\raisebox{-13pt}{10}]
};

\nextgroupplot[title={pulse\_detect}, yticklabels=\empty]
\begin{pgfonlayer}{axis background}
    \fill[BgGray] (axis cs:0.45,0) rectangle (axis cs:1.5,11.5);
    \fill[BgGray] (axis cs:2.5,0) rectangle (axis cs:3.55,11.5);
\end{pgfonlayer}
\addplot+[bar style] coordinates {(1,5) (2,2) (3,1)};
\addplot+[line style] coordinates {
    (1,0)  [\raisebox{5pt}{0}]
    (2,5)  [\raisebox{5pt}{\hspace{-6pt}5}]
    (3,10) [\raisebox{-13pt}{10}]
};

\nextgroupplot[title={multi\_pipe\_8bit}, yticklabels=\empty]
\begin{pgfonlayer}{axis background}
    \fill[BgGray] (axis cs:0.45,0) rectangle (axis cs:1.5,11.5);
    \fill[BgGray] (axis cs:2.5,0) rectangle (axis cs:3.55,11.5);
\end{pgfonlayer}
\addplot+[bar style] coordinates {(1,4) (2,1) (3,1)};
\addplot+[line style] coordinates {
    (1,6)  [\raisebox{5pt}{6}]
    (2,10) [\raisebox{-13pt}{10}]
    (3,10) [\raisebox{-13pt}{10}]
};

\nextgroupplot[title={div\_16bit}, yticklabels=\empty]
\begin{pgfonlayer}{axis background}
    \fill[BgGray] (axis cs:0.45,0) rectangle (axis cs:1.5,11.5);
    \fill[BgGray] (axis cs:2.5,0) rectangle (axis cs:3.55,11.5);
\end{pgfonlayer}
\addplot+[bar style] coordinates {(1,6) (2,1) (3,1)};
\addplot+[line style] coordinates {
    (1,2)  [\raisebox{5pt}{2}]
    (2,10) [\raisebox{-13pt}{10}]
    (3,10) [\raisebox{-13pt}{10}]
};

\end{groupplot}

\node[
    draw=none, fill=white,
    font=\footnotesize,
    inner xsep=6pt, inner ysep=3pt,
    anchor=north
] at ($(group c3r1.south) + (0, -0.55cm)$) {
    \begin{tabular}{rl@{\hspace{2em}}rl}
        \tikz[baseline=-0.3ex]{
            \filldraw[fill=BarBlue, draw=BarEdge, semithick] (0,0) rectangle (0.45,0.22);} & Distinct output behaviors 
         {\color{TextBlue}\(\boldsymbol{\downarrow}\)}
        &
        \tikz[baseline=-0.3ex]{
            \draw[LineRed, line width=1pt] (0,0.11) -- (0.6,0.11);
            \filldraw[LineRed, draw=white, thick] (0.3,0.11) circle (1.5pt);} & Officially verified candidates
             {\color{TextRed}\(\boldsymbol{\uparrow}\)}
    \end{tabular}
};

\end{tikzpicture}
\vspace{-0.1cm}
\caption{
Behavioral convergence and functional correctness on representative RTLLM hard tasks. Bars denote the number of distinct output behaviors, while lines denote the number of officially verified
candidates.
}
\label{fig:semantic_convergence}
\end{figure*}

We evaluate Layered Formal Guidance on four public benchmark sources covering different RTL tasks: specification-to-RTL generation from RTLLM~\cite{lu2024rtllm} and VerilogEval~\cite{VerilogEval}, RTL code generation tasks from the CVDP~\cite{cvdp}, and IP-level RTL generation from ChipVerilog~\cite{chipverilog}.We evaluate representative state-of-the-art RTL generation methods, including OriGen~\cite{2024origen}, RTLCoder,~\cite{Xie_2023} and CodeV-R1~\cite{zhuqimeng}, and general LLMs from the DeepSeek~\cite{deepseekai2026deepseekv4}, Qwen~\cite{qwen38}, and GPT families~\cite{openai_gpt53_2026}. 

During the generation process, we generate ten RTL candidates using a temperature of 0.7 and top-p of 0.9. For each task, the SVA set is generated once using the same LLM, and the same SVA set is used for all ten RTL candidates. CEX-guided refinement is limited to one iteration per candidate. Each candidate is considered correct only if it passes the official validation flow of the corresponding benchmark: the official testbenches for VerilogEval, RTLLM, and CVDP, and the official testbench and equivalence checking against the reference implementation for ChipVerilog. Functional correctness is reported using the standard \textit{pass@k} metric: 
\begin{equation}
    \text{pass@k} := \mathbb{E}_{\text{Problems}} \left[ 1 - \frac{\binom{n-c}{k}}{\binom{n}{k}} \right]
\end{equation}
where $n=10$ represents the total number of generated samples, $c$ is the number of candidates passing the official validation flow, and $k$ specifies the number of selected code samples. Simulation is performed using Synopsys VCS~\cite{synopsys_vcs}, while formal checking is performed using Yosys~\cite{yosys}. Local model inference is performed on a single NVIDIA A30 GPU, while API-hosted models such as GPT are accessed through their official APIs.

\subsection{Behavioral Divergence Analysis}

To analyze candidate diversity, we generated ten RTL candidates for each of five representative RTLLM tasks. We used behavior count to measure output diversity and pass count to measure verification success. As shown in Fig.~\ref{fig:semantic_convergence}, with the original natural-language specification (Orig.), each task produced 4 to 9 distinct behaviors, while only 0 to 6 candidates passed the official verification. This result shows that a natural-language specification may lead to different RTL behaviors and incorrect implementations. 

We then added explicit RTL requirements by augmenting the specification (Aug.). After applying it, the behavior count dropped to 1 to 3, and the official pass count increased to 5 to 10. We further applied CEX refinement (CEX) to resolve the mismatches among the specification, formal rules, and generated RTL. After one refinement round, all five tasks showed only one observed behavior, and all ten candidates passed the official verification. These results demonstrate that explicit RTL-level constraints narrow the generation space before generation, and CEX-guided refinement reduces the remaining inconsistencies after generation.

\begin{table}[h]
\centering
\caption{Quality of generated SVA properties and RTL generation results on FPU and OR1200 using DeepSeek-V4. Time denotes the average time per task.}
\label{tab:sva_quality_deepseek}
\renewcommand{\arraystretch}{1.12}
\setlength{\tabcolsep}{4pt}
\begin{tabularx}{\columnwidth}{
    >{\raggedright\arraybackslash}X
    c c c c
}
\toprule
\textbf{Metric} & \multicolumn{2}{c}{\textbf{FPU}} & \multicolumn{2}{c}{\textbf{OR1200}} \\
\midrule
Average \#SVA / Task     & \multicolumn{2}{c}{18.3}   & \multicolumn{2}{c}{17.3} \\
Syntactically Valid Rate & \multicolumn{2}{c}{85.8\%} & \multicolumn{2}{c}{91.0\%} \\
Non-vacuity Rate         & \multicolumn{2}{c}{91.9\%} & \multicolumn{2}{c}{88.8\%} \\
\midrule
\textbf{Setting} & \textbf{Pass@1} & \textbf{Time (s)} & \textbf{Pass@1} & \textbf{Time (s)} \\
\midrule
Original Spec             & 5.6\%  & 94.8 & 12.2\% & 88.5 \\
Original Spec + CEX       & 5.6\%  & 142.6 & 18.1\% & 137.5 \\
Augmented Spec            & 11.1\% & 102.5 & 29.3\% & 91.6 \\
Augmented Spec + CEX      & 22.2\% & 165.1 & 40.2\% & 145.9 \\
\bottomrule
\end{tabularx}
\end{table}

\subsection{Effectiveness of SVA Properties}
\label{subsec:sva_quality}

Table~\ref{tab:sva_quality_deepseek} reports the quality of model-generated SVA properties and their impact on RTL generation using DeepSeek-V4-Flash. We evaluate SVA quality using two metrics: syntactic validity, the fraction of generated properties accepted by the formal tool, and non-vacuity, the fraction of executable properties triggered by at least one formal run or simulation trace under the task setup. For downstream RTL generation we report Pass@1 and the average time per task, including LLM inference and formal checking.

The generated properties achieve high validity and non-vacuity rates. The syntactically valid rates are 85.8\% on FPU and 91.0\% on OR1200, while the non-vacuity rates are 91.9\% and 88.8\%, respectively. These properties also improve RTL correctness when used as guidance. Specification augmentation raises Pass@1 from 5.6\% to 11.1\% on FPU and from 12.2\% to 29.3\% on OR1200. CEX-guided refinement is considerably more effective when applied on the augmented specification. On OR1200, it improves Pass@1 from 12.2\% to 18.1\% with the original specification and from 29.3\% to 40.2\% with the augmented specification. These results show that model-generated SVA properties provide effective semantic constraints for RTL generation, particularly on control- and state-intensive designs.

\subsection{Correctness Comparison}
\definecolor{lightpurple}{RGB}{245,240,252}

\newcommand{\pcell}[1]{\cellcolor{lightpurple}#1}

\begin{table*}[t]
\centering
\footnotesize

\renewcommand{\arraystretch}{1.15}
\setlength{\tabcolsep}{4.5pt}
\setlength{\aboverulesep}{0.4ex}
\setlength{\belowrulesep}{0.4ex}

\caption{Pass@$k$ results on the FPU (9 tasks), OR1200 (38 tasks), and CVDP (76 tasks) benchmark suites.}
\label{tab:chipverilog_result}

\vspace{-0.5mm}

\begin{tabularx}{\textwidth}{
@{}
l
>{\raggedright\arraybackslash}X
c c c
c c c
c c c
@{}
}
\toprule

\multirow{2}{*}{\textbf{Model}} &
\multirow{2}{*}{\textbf{Setting}} &
\multicolumn{3}{c}{\textbf{FPU}} &
\multicolumn{3}{c}{\textbf{OR1200}} &
\multicolumn{3}{c}{\textbf{CVDP}} \\

\cmidrule(lr){3-5}
\cmidrule(lr){6-8}
\cmidrule(lr){9-11}

& &
pass@1 &
pass@5 &
pass@10 &
pass@1 &
pass@5 &
pass@10 &
pass@1 &
pass@5 &
pass@10 \\

\midrule

\multirow{3}{*}{Qwen 3.6}
& Base model
& 2.22 & 11.11 & 22.22
& 29.48 & 40.97 & 42.11
& 31.15 & 43.23 & 50.00 \\

& Base model + SVA
& 13.33 & 33.95 & 44.44
& 34.53 & 45.27 & 50.53
& 35.38 & 60.39 & 69.23 \\

& Base model + SVA + CEX
& \pcell{17.78} & \pcell{43.25} & \pcell{55.56}
& \pcell{42.52} & \pcell{52.41} & \pcell{54.74}
& \pcell{36.41} & \pcell{61.85} & \pcell{70.51} \\

\midrule

\multirow{3}{*}{DeepSeek-Coder}
& Base model
& 0.00 & 0.00 & 0.00
& 0.45 & 2.25 & 4.51
& 13.64 & 21.78 & 24.24 \\

& Base model + SVA
& 3.33 & 10.19 & 11.11
& 6.32 & 20.16 & 27.07
& 27.56 & 36.77 & 39.74 \\

& Base model + SVA + CEX
& \pcell{4.45} & \pcell{10.85} & \pcell{11.11}
& \pcell{11.73} & \pcell{33.73} & \pcell{40.60}
& \pcell{32.56} & \pcell{40.61} & \pcell{42.31} \\

\midrule

\multirow{3}{*}{DeepSeek-V4}
& Base model
& 5.55 & 11.07 & 11.11
& 12.18 & 34.98 & 40.60
& 49.36 & 55.97 & 56.41 \\

& Base model + SVA
& 11.11 & 25.31 & 33.33
& 29.32 & 54.99 & 58.65
& 50.64 & 57.25 & 57.69 \\

& Base model + SVA + CEX
& \pcell{22.22} & \pcell{46.34} & \pcell{55.56}
& \pcell{40.15} & \pcell{57.52} & \pcell{58.65}
& \pcell{55.64} & \pcell{63.65} & \pcell{64.10} \\

\midrule

\multirow{3}{*}{GPT-4o}
& Base model
& 4.45 & 10.85 & 11.11
& 11.28 & 31.47 & 36.09
& 44.49 & 57.97 & 61.54 \\

& Base model + SVA
& 7.78 & 16.67 & 22.22
& 26.17 & 49.71 & 54.13
& 48.85 & 63.10 & 67.95 \\

& Base model + SVA + CEX
& \pcell{11.11} & \pcell{26.85} & \pcell{33.33}
& \pcell{32.93} & \pcell{54.60} & \pcell{58.65}
& \pcell{50.00} & \pcell{63.74} & \pcell{67.95} \\

\midrule

\multirow{3}{*}{GPT-5.4}
& Base model
& 12.22 & 28.40 & 33.33
& 31.58 & 50.46 & 54.13
& 56.94 & 63.70 & 65.28 \\

& Base model + SVA
& 22.22 & 54.32 & 66.67
& 45.94 & 69.39 & 73.68
& 72.78 & 77.78 & 79.17 \\

& Base model + SVA + CEX
& \pcell{43.33} & \pcell{78.13} & \pcell{88.89}
& \pcell{62.63} & \pcell{72.08} & \pcell{73.68}
& \pcell{75.83} & \pcell{80.56} & \pcell{81.94} \\

\midrule

\multicolumn{2}{l}{Avg. gain of SVA}
& +6.67 & +15.80 & +20.00
& +11.46 & +15.88 & +17.32
& +7.75 & +8.52 & +9.21 \\

\multicolumn{2}{l}{Avg. gain of SVA + CEX}
& +14.89 & +28.80 & +33.34
& +21.00 & +22.04 & +21.78
& +10.79 & +11.54 & +11.82 \\

\bottomrule
\end{tabularx}

\vspace{-1mm}
\end{table*}
Table~\ref{tab:chipverilog_result} summarizes the pass@k results of five LLMs on the FPU, OR1200, and CVDP benchmarks under three settings: the base model, the base model with SVA-based specification augmentation (+SVA), and the full framework with CEX-guided refinement (+SVA+CEX). Layered Formal Guidance improves all five models across the three benchmarks. Specification augmentation improves the average pass@1 by 6.67 pp on FPU, 11.46 pp on OR1200, and 7.75 pp on CVDP. CEX refinement adds another 8.22 pp, 9.54 pp, and 3.05 pp, respectively. These results show that the two stages provide complementary gains. Specification augmentation restricts semantic ambiguity before RTL generation, while CEX refinement addresses the remaining inconsistencies after generation.

The improvement is consistent across benchmark suites. The average pass@1 increases from 4.89\% to 19.78\% on FPU, from 16.99\% to 37.99\% on OR1200, and from 39.3\% to 50.09\% on CVDP. The gains are larger on the two IP-level benchmarks, reaching 14.89 pp on FPU and 21.00 pp on OR1200. These results show that our method remains effective on larger and more complex RTL designs. CVDP also shows a 10.79 pp gain despite its higher base-model accuracy. The magnitude of refinement gain also depends on the base model. On FPU, GPT-5.4 gains 21.11 pp from the CEX stage, whereas DeepSeek-Coder gains only 1.12 pp. While the formal tool supplies detailed counterexample traces, the performance gains depend on the model's ability to analyze these execution traces and reason through complex logic to generate a correct repair. 

\subsection{Generalizability Across LLM Architectures}
\begin{table}[t]
\centering
\caption{Compatibility and performance gains of Layered Formal Guidance across diverse LLMs and specialized RTL generation systems. }
\label{tab:compatibility_results}

\scriptsize
\setlength{\tabcolsep}{1.8pt}
\renewcommand{\arraystretch}{0.96}

\begin{tabular}{@{}llcccccc@{}}
\toprule
\textbf{Model}
& \textbf{Setting}
& \multicolumn{2}{c}{\textbf{VerilogEvalMachine}}
& \multicolumn{2}{c}{\textbf{VerilogEvalHuman}}
& \multicolumn{2}{c}{\textbf{RTLLM 2.0}} \\
\cmidrule(lr){3-4}
\cmidrule(lr){5-6}
\cmidrule(lr){7-8}
&
& pass@1 & pass@5
& pass@1 & pass@5
& pass@1 & pass@5 \\
\midrule

RTLCoder
& Base 7B
& 61.20 & 76.50
& 41.60 & 50.10
& 43.50 & 48.10 \\

& \textbf{+SVA+CEX}
& 65.26 & 79.80
& 49.04 & 57.17
& 62.90 & 66.27 \\

\addlinespace[1pt]
OriGen
& Base 7B
& 74.10 & 82.40
& 54.40 & 60.10
& 50.90 & 60.90 \\

& \textbf{+SVA+CEX}
& 78.23 & 85.55
& 61.90 & 67.20
& 69.10 & 74.40 \\

\addlinespace[1pt]

CodeV-R1
& Base 8B
& 76.50 & 84.10
& 69.90 & 79.30
& 72.90 & 86.10 \\

& \textbf{+SVA+CEX}
& 80.14 & 87.25
& 77.14 & 86.35
& 89.70 & 97.21 \\

\addlinespace[1pt]

Qwen3.6 
& Base 35B
& 74.13 & 80.05
& 55.32 & 67.27
& 52.34 & 60.70 \\

& \textbf{+SVA+CEX}
& 81.89 & 90.87
& 69.62 & 85.29
& 75.74 & 87.18 \\

\addlinespace[1pt]

DeepSeek-V4
& Base 284B
& 76.43 & 83.07
& 65.19 & 75.62
& 52.98 & 70.54 \\

& \textbf{+SVA+CEX}
& 79.86 & 91.47
& 71.28 & 87.86
& 70.21 & 90.68 \\

\midrule

\multicolumn{2}{l}{\textbf{Mean paired gain (pp)}}
& \textbf{+4.60} & \textbf{+5.76}
& \textbf{+8.51} & \textbf{+10.30}
& \textbf{+19.01} & \textbf{+17.88} \\

\bottomrule
\end{tabular}
\end{table}
We evaluate our methods on VerilogEval and RTLLM 2.0 with three RTL-specific methods, OriGen, RTLCoder, and CodeV-R1, and two general-purpose LLMs, Qwen3.6 and DeepSeek-V4. For each method, we keep its original generation process and use the same base model for SVA generation, RTL generation, and CEX analysis. Table~\ref{tab:compatibility_results} shows consistent improvements across all five methods. The average pass@1 gains are 4.60 pp on VerilogEval-Machine, 8.51 pp on VerilogEval-Human, and 19.01 pp on RTLLM 2.0. The improvement also holds for state-of-the-art domain-specific models. CodeV-R1 improves from 72.90\% to 89.70\% on RTLLM 2.0, while OriGen improves by 7.50 pp on VerilogEval-Human and 18.20 pp on RTLLM 2.0. These results show that Layered Formal Guidance is model-agnostic and seamlessly integrates with existing feedback-driven and post-generation repair methods. 

\section{Conclusion}
\label{sec:conclusion}

This paper presented Layered Formal Guidance, a framework designed to bridge the semantic gap between high-level natural-language specifications and implementation-critical RTL details. The key idea is to apply formal guidance in layer-specific forms: structured constraints before generation, assertion-based correctness checks after generation, and counterexample-based refinement signals after verification. This layered use of formal properties narrows the LLM's generation space, verifies generated RTL against formal assertions, and identifies whether the RTL, the specification, or the properties should be refined. Across RTLLM hard tasks and OpenCores-derived IP designs, Layered Formal Guidance improved functional correctness and behavioral convergence across multiple LLMs. These results show that formal properties can serve not only as post-generation checkers, but also as active guidance for more reliable RTL generation. 

\bibliographystyle{IEEEtran}
\bibliography{references}

\end{document}